\documentclass{IEEEtran}
\usepackage[T1]{fontenc}
\usepackage[utf8]{inputenc}
\usepackage{array}
\usepackage{verbatim}
\usepackage{multirow}
\usepackage{amsmath}
\usepackage{graphicx}
\usepackage{xcolor}
\definecolor{darkcyan}{rgb}{0.0, 0.35, 0.55}

\usepackage[colorlinks=true, linkcolor=darkcyan, citecolor=black, urlcolor=black, pdfborder={0 0 0}]{hyperref}    % arxiv

\makeatletter

\providecommand{\tabularnewline}{\\}
\let\oldforeign@language\foreign@language
\DeclareRobustCommand{\foreign@language}[1]{%
  \lowercase{\oldforeign@language{#1}}}
\usepackage{amsfonts}\usepackage{algorithmic}
\usepackage{array}
\usepackage{textcomp}
\usepackage{stfloats}
\usepackage{url}
\usepackage{verbatim}
\usepackage{soul}\usepackage{threeparttable}% for table notes
\usepackage{algorithmic}

\def\BibTeX{{\rm B\kern-.05em{\sc i\kern-.025em b}\kern-.08em
    T\kern-.1667em\lower.7ex\hbox{E}\kern-.125emX}}
\usepackage{balance}
\usepackage{cite}

\makeatother

\usepackage{fancyhdr}

\fancypagestyle{firstpage}{%
  \fancyhf{} % clear header/footer
  
  \fancyhead[L]{\parbox{0.99\textwidth}{\scriptsize
    \copyright~2024 IEEE. Personal use of this material is permitted.  Permission from IEEE must be obtained for all other uses, in any current or future media, including reprinting/republishing this material for advertising or promotional purposes, creating new collective works, for resale or redistribution to servers or lists, or reuse of any copyrighted component of this work in other works. 
    DOI: 10.1109/LSP.2024.3483008}}%
}

\begin{document}
\title{Order Estimation of Linear-Phase FIR Filters for DAC Equalization
in Multiple Nyquist Bands}
\author{Deijany Rodriguez Linares, \textit{Graduate Student Member, IEEE},\thanks{Deijany Rodriguez Linares and Håkan Johansson are with the Department
of Electrical Engineering, Linköping University, 58183 Linköping,
Sweden (e-mail: deijany.rodriguez.linares@liu.se; hakan.johansson@liu.se).} \and Håkan Johansson, \textit{Senior Member, IEEE}, and \\Yinan
Wang, \textit{Member, IEEE} \thanks{Yinan Wang is with the College of Electronic Science and Technology,
National University of Defense Technology, Changsha, Hunan 410073
China (e-mail: wangyinan@nudt.edu.cn).} \\ \thanks{\textit{Corresponding authors: D. R. Linares, H. Johansson, and Y. Wang.} } }
\markboth{}{}
\maketitle
\thispagestyle{firstpage} 

\begin{abstract}
This letter considers the design and properties of linear-phase finite-length impulse
response (FIR) filters for equalization of the frequency responses
of digital-to-analog converters (DACs). The letter derives estimates
for the filter orders required, as functions of the bandwidth and
equalization accuracy, for four DAC pulses that are used in DACs in
multiple Nyquist bands. The estimates are derived through a large set of minimax-optimal equalizers
and the use of symbolic regression followed by minimax-optimal curve
fitting for further enhancement. Design examples demonstrate the accuracy of the proposed estimates. In addition, the letter discusses
the appropriateness of the four types of linear-phase FIR filters,
for the different equalizer cases, as well as the corresponding properties
of the equalized systems.
\end{abstract}


\section{Introduction\label{sec:Introduction}}

Digital-to-analog converters (DACs) constitute fundamental components
as conversions between digital and analog signals are needed in virtually
all signal processing systems \cite{Johansson_14_0}. A DAC makes
use of a simple pulse (see Fig. 1 in Section \ref{sec:DAC-Pulses-and})
which introduces linear distortion, i.e., a non-constant frequency
response in the signal band of interest. To reduce the in-band distortion,
equalization is required and preferably carried out in the digital
domain before the DAC. Further, in wide-band systems with high sampling
rates, the equalization employs finite-length impulse response
(FIR) filters, as infinite-length impulse response (IIR) filters impose
upper bounds on the sampling rate \cite{Renfors_81,Wanhammar_13}.
In the past, several papers have considered the design of FIR equalizers
but their focus has been on the first Nyquist band (NB) and the conventional
non-return-to-zero (NRTZ) {[}a.k.a. zero-order hold (ZOH){]} DAC pulse
\cite{Samueli_1988,Vankka_2002}. For higher NBs, other pulses can
be more appropriate (see Section \ref{sec:DAC-Pulses-and}). In particular,
we consider three additional DAC pulses that are used in multi-mode
DACs for multiple NBs \cite{Overhoff_2018}, viz. return-to-zero (RTZ),
return-to-complement (RTC) {[}a.k.a. mixed mode or radio frequency
(RF){]}, and return-to-complement-to-zero (RTCZ) {[}a.k.a. radio frequency
return-to-zero (RFZ){]}. Recently, \cite{Lan19} considered the RTZ
and RTC cases as well, but the design was based on frequency sampling
which does not result in minimax-optimal filters. Further, only the
magnitude response was equalized in \cite{Lan19} but not the phase
response in the RTC and RTCZ cases because of an additional imaginary
unit $j$ in the DAC frequency response (see Section \ref{sec:DAC-Pulses-and}).
Furthermore, in the literature, there is a lack of detailed analysis
regarding filter order versus bandwidth and equalization accuracy.
In view of the above, the contribution of this letter is as follows.

The letter derives equations that estimate the minimal filter order required to meet
a specified bandwidth and equalization accuracy, for the four pulses
mentioned above and for the corresponding appropriate NBs (see Table
\ref{tab:Filter properties} in Section \ref{sec:Equalizer-Classes}). The proposed estimates are derived through a large set of minimax-optimal
equalizers and the use of symbolic regression and minimax-optimal
curve fitting for further improvement. Design examples demonstrate the accuracy of the proposed estimates.
Moreover, the letter discusses the appropriateness of the four
types of linear-phase FIR filters, for the different equalizer cases,
as well as the corresponding properties of the equalized systems.
It is noted that order estimates for linear-phase FIR filters have appeared
before in the literature, but only for frequency-selective filters
\cite{Ichige2000,Lai_2008}, differentiators \cite{Sheikh2011,Wang_2023},
and analog-to-digital converter (ADC) bandwidth extension \cite{Wang_2019}.
Hence, the existing order estimates are not applicable for FIR equalizers
in different NBs.

After this introduction, Section \ref{sec:DAC-Pulses-and} considers
the four DAC pulses and the basics of linear-phase FIR filters. Section
\ref{sec:Equalizer-Classes} discusses the different equalizers and
properties whereas Section \ref{sec:Proposed-Filter-Order-Estimates}
derives the proposed order estimates. Section \ref{sec:Results} provides
design results, verifying the accuracy of the proposed estimates,
whereas Section \ref{sec:Conclusion} concludes the letter.

\section{DAC Pulses and Linear-Phase FIR Filters\label{sec:DAC-Pulses-and}}

\begin{figure}[t!]
\centering \includegraphics[scale=0.8]{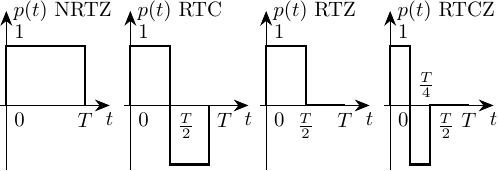} \caption{DAC pulses.}
\label{impulse_response}
\end{figure}

\begin{figure}[t!]
\centering 
 \includegraphics[scale=0.8]{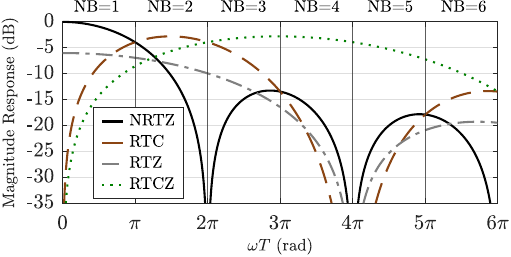} \caption{Magnitude responses of the pulses.}
 \label{magnitude_response}
\end{figure}

\textit{DAC Pulses: }We consider the four DAC pulses NRTZ, RTZ, RTC,
and RTCZ, depicted in Fig. \ref{impulse_response}.
The frequency responses (Fourier transforms) of these pulses can be
written as %%(utilizing Euler's formula and trigonometric identities)
\begin{align}
P(j\omega) & =\begin{cases}
T\times\frac{\sin(\frac{\omega T}{2})}{\omega T/2}e^{-j\frac{\omega T}{2}}, & \text{NRTZ}\\
T\times\frac{1}{2}\frac{\sin(\frac{\omega T}{4})}{\omega T/4}e^{-j\frac{\omega T}{4}}, & \text{RTZ}\\
T\times j\frac{1-\cos(\frac{\omega T}{2})}{\omega T/2}e^{-j\frac{\omega T}{2}}, & \text{RTC}\\
T\times\frac{1}{2}j\frac{1-\cos(\frac{\omega T}{4})}{\omega T/4}e^{-j\frac{\omega T}{4}}, & \text{RTCZ}
\end{cases}\label{eq:freqResp}
\end{align}
where $T$ denotes the sampling period, $\omega$ denotes the angular
frequency (rad/s), and $\omega T$ denotes the digital-signal normalized
frequency (rad). Figure \ref{magnitude_response} plots the magnitude
responses (modulus of the Fourier transforms) across the first six
NBs. It is seen that the NRTZ pulse has a relatively flat magnitude
response in the first NB but large droops in the higher NBs due to
a zero at the sampling frequency ($\omega T=2\pi$) and multiples
thereof. Therefore, the other pulses have been proposed for the higher
NBs as they offer flatter magnitude responses in those bands and lower
out-of-band gains which are easier to suppress by the analog filters
that follow the DAC \cite{Overhoff_2018}. As seen in Fig. \ref{magnitude_response},
the RTZ pulse has a relatively flat response over the first three
NBs whereas the RTC pulse is appropriate only for the second and third
NB. The RTCZ pulse has a relatively flat response in all NBs except
for the first one.

\textsl{Linear-Phase FIR Filters:} The frequency response of a causal
$N$th-order FIR filter is 
\begin{equation}
H(e^{j\omega T})=\sum_{n=0}^{N}h(n)e^{-j\omega Tn},\label{eq:generalFIR}
\end{equation}
where $h(n)$ denotes the impulse response which is assumed to be
real-valued throughout this letter. For a linear-phase FIR filter,
the impulse response is either symmetric (Type I and II, for even
and odd $N$, respectively) or antisymmetric (Type III and IV, for
even and odd $N$, respectively). In this case, the frequency response
can be written as 
\begin{align}
H(e^{j\omega T}) & =\begin{cases}
e^{-j\omega TN/2}H_{R}(\omega T), & \text{Type I, II,}\\
je^{-j\omega TN/2}H_{R}(\omega T), & \text{Type III, IV,}
\end{cases}\label{eq:filter_type}
\end{align}
where $H_{R}(\omega T)$ denotes the real-valued zero-phase frequency
response \cite{Saramaki_93,Jackson_96,Wanhammar_13}. Further, Types
II, III, and IV have restrictions due to structural zeros at $\omega T=0$
and/or $\omega T=\pi$, as seen in Table \ref{tab:features} which
summarizes some of the properties of linear-phase FIR filters. It is noted that the use of linear-phase
FIR filters preserves the linear phase in the equalized system. Further, linear-phase FIR filters
require only about half as many multipliers, due to the impulse response symmetry or antisymmetry, which leads
to a lower implementation complexity. The number of multipliers may be further reduced by using fewer distinct impulse response values, but to still meet a given specification, it will generally also result in a filter order above the minimal filter order. Such a technique was considered in \cite{Nassralla_2021} for frequency selective filters. Applying that technique to the equalizers reveals that multiplier savings seem feasible for some filter-type and NB combinations, but not for all. However, to fully investigate the potential benefits of using the technique in \cite{Nassralla_2021} for the equalizers, further comprehensive studies are needed which is beyond the scope of this letter whose focus is to derive estimates of the minimal filter orders required, as functions of the bandwidth and estimation accuracy, which provide a starting point for such studies.

\begin{table}[t]
\centering{}\centering \caption{Properties of Linear-Phase FIR Filters.}
\label{tab:features} %
\begin{tabular}{|c|c|c|c|c|}
\hline 
Type & Order & Imp. resp. & Multipliers & Structural zeros\tabularnewline
\hline 
I & Even & Sym. & $N/2+1$ & None\tabularnewline
\hline 
II & Odd & Sym. & $(N+1)/2$ & One, at $\omega T=\pi$\tabularnewline
\hline 
III & Even & Antisym. & $N/2$ & Two, at $\omega T=0\textnormal{ and }\pi$\tabularnewline
\hline 
IV & Odd & Antisym. & $(N+1)/2$ & One, at $\omega T=0$\tabularnewline
\hline 
\end{tabular}
\end{table}

\begin{table}
	\caption{Equalizers and Delay of the Linear-Phase Equalized System. Here, INT
		Represents an Integer Given by \eqref{eq:INT}. }
	
	\begin{centering}
		\begin{tabular}{|c|c|c|c|}
			\hline 
			Pulse & NB & Type & Delay, $K$\tabularnewline
			\hline 
			\hline 
			\multirow{2}{*}{NRTZ} & \multirow{2}{*}{1} & \multicolumn{1}{c|}{I} & $\text{INT}+1/2$\tabularnewline
			\cline{3-4} \cline{4-4} 
			&  & II & $\text{INT}$\tabularnewline
			\hline 
			\multirow{2}{*}{RTZ} & \multirow{2}{*}{1--3} & I & $\text{INT}+1/4$\tabularnewline
			\cline{3-4} \cline{4-4} 
			&  & II & $\text{INT}-1/4$\tabularnewline
			\hline 
			\multirow{2}{*}{RTC} & \multirow{2}{*}{2, 3} & III & $\text{INT}+1/2$\tabularnewline
			\cline{3-4} \cline{4-4} 
			&  & IV & $\text{INT}$\tabularnewline
			\hline 
			\multirow{2}{*}{RTCZ} & \multirow{2}{*}{2--6} & III & $\text{INT}+1/4$\tabularnewline
			\cline{3-4} \cline{4-4} 
			&  & IV & $\text{INT}-1/4$\tabularnewline
			\hline 
		\end{tabular}
		\par\end{centering}
	\label{tab:Filter properties}
\end{table}

\section{Equalizers and Their Properties \label{sec:Equalizer-Classes}}

Based on the equations and discussions in Section \ref{sec:DAC-Pulses-and},
the equalizers summarized in Table \ref{tab:Filter properties} are
considered. It is noted that only Type III and Type IV filters can
be used for equalization in the RTC and RTCZ cases, due to the imaginary
unit $j$ in the DAC frequency response seen in \eqref{eq:freqResp}
and in the filter frequency response seen in \eqref{eq:filter_type}.
Likewise, only Type I and Type II filters can be used for the NRTZ
and RTZ cases as $j$ is then not present in the frequency responses.
Further, the term $\exp(-j\omega T/2)$ for NRTZ and RTC, and the
term $\exp(-j\omega T/4)$ for RTZ and RTCZ, correspond to half-sample
and quarter-sample delays, respectively. These terms are normally
excluded in the equalizer designs (as in this letter), as they can be either ignored or
taken care of by other means in the overall systems where the DACs
are used\footnote{Using Type II or Type IV linear-phase FIR filters, which also have
half-sample delays, the half-sample delay for NRTZ and RTC is automatically
compensated, so that an integer-sample delay remains after equalization.
Using Type I or Type III linear-phase FIR filters, the half-sample
delay will remain after the equalization. For all four filter types,
a quarter-sample delay remains after the equalization in the RTZ and
RTCZ cases.}. Table \ref{tab:Filter properties} also includes the delays after
equalization, where INT represents an integer given by 
\begin{align}
\text{INT} & =\begin{cases}
N/2, & N\text{ even, Type I and III}\\
(N+1)/2, & N\text{ odd, Type II and IV.}
\end{cases}\label{eq:INT}
\end{align}
Further, as Type I linear-phase filters have no structural zeros,
they can theoretically cover a whole NB for NRTZ and RTZ. However,
the filter order tends to be very high and grows rapidly when the
bandwidth approaches a full NB. Therefore, a don't-care band is normally
adopted near $\pi$ when operating in the first NB. Likewise, it is customary to have
two don't-care bands in higher NBs and assume that the frequency band is centered
in the middle. Consequently, we assume that $\omega T\in\Omega$ covers
frequencies in the NBs according to\footnote{As real-valued impulse responses are assumed, it suffices to consider
positive frequencies in the design.} 
\begin{align}
\Omega & =\begin{cases}
\bigr[0,B\bigr], & NB=1\\
\bigr[(NB-1/2)\pi-\frac{B}{2},\,(NB-1/2)\pi+\frac{B}{2}\bigr], & NB>1
\end{cases}\label{eq:bandwidth}
\end{align}
where $B$ denotes the digital-signal bandwidth satisfying $0<B<\pi$
(see Fig. \ref{magnitude_response}). To avoid a high filter order, the bandwidth $B$ should not be too close to $\pi$,
as the results in Section \ref{sec:Results} will show.

\section{Proposed Filter-Order Estimates\label{sec:Proposed-Filter-Order-Estimates}}

The proposed filter-order estimates are derived in three steps as
described below, under the assumption that the equalizers are designed
in the minimax sense. For a given pulse, NB, filter type and
order $N$, the equalizer design amounts to optimizing the impulse
response values $h(n)$ to minimize the equalization accuracy parameter (approximation error) $\delta_{N}$, according to 
\begin{equation}
\begin{array}{ll}
\text{minimize } & \delta_{N}\\
\text{subject to} & |E(j\omega T)|\leq\delta_{N}\quad\forall\omega T\in\Omega,
\end{array}\label{eq:optim-problem}
\end{equation}
where $\Omega$ is given by \eqref{eq:bandwidth} and the error function
$E(j\omega T)$ is
\begin{equation}
E(j\omega T)=H(e^{j\omega T})P(j\omega)/T-e^{-j\omega TK},\label{eq:error-function}
\end{equation}
since $T$ in \eqref{eq:freqResp} should be excluded in the equalization\footnote{Recall that $P(j\omega)=T$ for $|\omega|<\pi/T$ and $P(j\omega)=0$
for $|\omega|>\pi/T$ corresponds to perfect reconstruction \cite{Johansson_14_0}.}. Further, $K$ is the delay given in Table \ref{tab:Filter properties}
and \eqref{eq:INT}. As the problem in \eqref{eq:optim-problem} is
convex, the designed filter is globally optimal in the minimax sense
\cite{Saramaki_93,Ching-Yih_1994,Jackson_96,Wanhammar_13}. For the
design, one can use any solver for convex problems. We have used the
fast Parks-McClellan-Rabiner algorithm\footnote{The delay term $e^{-j\omega TK}$ in \eqref{eq:error-function} is
then factored out and the optimization problem is restated in terms
of the real-valued zero-phase functions.} \cite{McClellan_73} which is tailored for linear-phase FIR filters,
implemented in e.g. the function \textit{firpm} in Matlab, and works
well for all equalizers and orders (up to some 300) under consideration
in Section \ref{sec:Results}\footnote{For higher-order filters, other design methods may be more robust,
like the one in \cite{Zahradnik_2020} developed for frequency selective
Type I filters.}.

\textbf{Step 1:} Determine a set of desired bandwidths $B\in[B_{1},B_{2}]$
where $B_{1}$ and $B_{2}$ represent the most narrow and wide bandwidths,
respectively, and a set of desired equalization accuracy $\delta\in[\delta_{1},\delta_{2}]$
where $\delta_{1}$ and $\delta_{2}$ represent the smallest and largest
accuracy, respectively. For each pair of values ($B,\delta$) in these
sets, determine the minimal filter order, $N_{\text{min}}(B,\delta)$,
required to meet the specification. Here, $N_{\text{min}}(B,\delta)$
is found by solving \eqref{eq:optim-problem} several times with increasing
filter order until the specification is satisfied, i.e., when $\delta_{N}\leq\delta$.
As the problem in \eqref{eq:optim-problem} is convex, $N_{\text{min}}(B,\delta)$
is the minimal order required to achieve an accuracy of $\delta$
for the bandwidth $B$.

\textbf{Step 2:} Based on all values of $N_{\text{min}}(B,\delta)$
obtained in Step 1, use a symbolic regression-based algorithm \cite{Chen17,deFranca2022},
with $\pi-B$ and $\log_{10}(\delta)$ as inputs (based on existing
formulas for order estimation of linear-phase FIR filters for other
applications), with the aim of approximating $N_{\text{min}}(B,\delta)$
as well as possible. Symbolic regression uses genetic programming
in its search for the best-fit mathematical expression. Here, we assume
at most three terms in the order estimate to keep it reasonably simple,
and obtain the best basic expression as 
\begin{equation}
N_{\text{est}}(B,\delta)=c+\frac{a\log_{10}(\delta)}{(\pi-B)}+\frac{b[\log_{10}(\delta)]^{2}}{(\pi-B)^{2}},\label{eq:Nest-initial}
\end{equation}
where $a$, $b$, and $c$ are parameters that take on different values
for the different equalizer types and pulses. 

\textbf{Step 3:} In the expression obtained in \eqref{eq:Nest-initial},
the powers of $\log_{10}(\delta)$ and $(\pi-B)$ were restricted
to take on integer values. To relax this restriction, and improve
the estimates, the powers of $\log_{10}(\delta)$ and $(\pi-B)$ can
here be any real numbers, and we use curve fitting to find the optimal
parameters. For this purpose, we use minimax optimization to minimize
the maximal distance between the estimated and minimal filter
orders over all specifications. This problem is stated as 
\begin{equation}
\begin{array}{ll}
\text{minimize } & \varepsilon\\
\text{subject to} & |N_{\text{est}}(B,\delta)-N_{\text{min}}(B,\delta)|\leq\varepsilon
\end{array}\label{eq:optim-problem-curvefitting}
\end{equation}
for all $B\in[B_{1},B_{2}]$ and $\delta\in[\delta_{1},\delta_{2}]$
where 
\begin{equation}
N_{\text{est}}(B,\delta)=c+\frac{a_{1}[\log_{10}(a_{2}\delta)]^{a_{3}}}{(\pi-B)^{a_{4}}}+\frac{b_{1}[\log_{10}(b_{2}\delta)]^{b_{3}}}{(\pi-B)^{b_{4}}}.\label{eq:Nest}
\end{equation}
To solve this problem, we use the function \textit{fminimax} in Matlab
with initial parameter values obtained ($a$, $b$, and $c$) and
used (values one and two) in Step 2, i.e., $c$, $a_{1}=a$, and $b_{1}=b$,
and $a_{2}=a_{3}=a_{4}=b_{2}=1$ and $b_{3}=b_{4}=2$. We also constrain
the parameter values so that the second and third terms in \eqref{eq:Nest}
are not interchanged during the optimization. 

As the optimization problem in \eqref{eq:optim-problem-curvefitting}
is non-convex, there is no guarantee that the locally optimal solution
is globally optimal. What can be stated though, is that the proposed
three-step optimization procedure gives accurate filter order estimates, as will
be verified in Section \ref{sec:Results}. It is also noted that existing
filter order estimates, for other applications, correspond to non-convex
optimization problems as well, so this is not unique for the proposed
estimates.

\section{Results\label{sec:Results}}

The results are obtained via the three-step procedure proposed in
Section \ref{sec:Proposed-Filter-Order-Estimates} for a bandwidth
$B$ and approximation error $\delta$, ranging from $B_{1}=0.04\pi$
to $B_{2}=0.96\pi$, and from $\delta_{1}=10^{-5}$ to $\delta_{2}=10^{-1}$,
respectively\footnote{For $B$ ($\delta$), we have used $150$ ($50$) values linearly (logarithmically) spaced between $B_{1}$
and $B_{2}$ ($\delta_{1}$ and $\delta_{2}$).}.

\begin{table*}[htbp]
\centering \caption{Parameters for the Proposed Order Estimate in \eqref{eq:Nest}.}
\label{tab:parameters}%
\begin{tabular}{|c|c|c|c|c|c|c|c|c|c|c|c|c|}
\hline 
Pulse & NB & Type & $a_{1}$ & $a_{2}$ & $a_{3}$ & $a_{4}$ & $b_{1}$ & $b_{2}$ & $b_{3}$ & $b_{4}$ & $c$ & $\varepsilon_{\text{max}}$\tabularnewline
\hline 
\hline 
NRTZ & 1 & I & -1.8860 & 1.6994 & 1.3380 & 1.2150 & 1.1523 & 3.1189 & 0.8861 & 1.5646 & 1.6565 & 2.07\tabularnewline
\hline 
NRTZ & 1 & II & -2.5636 & 0.0419 & 1.1841 & 1.0280 & 1.3290 & 5.3295 & -0.0435 & 1.5107 & -2.1359 & 2.15\tabularnewline
\hline 
RTC & 2 & III & -6.5450 & 0.2535 & 1.1020 & 1.0592 & 0.7124 & 0.0144 & 0.5291 & 1.6930 & -3.3270 & 4.26\tabularnewline
\hline 
RTC & 2 & IV & -6.1858 & 0.1881 & 1.1211 & 1.0690 & 0.6163 & 0.0004 & 0.6623 & 1.6317 & -3.7629 & 3.42\tabularnewline
\hline 
RTC & 3 & III & -7.3009 & 0.3264 & 1.0470 & 1.0729 & 0.6768 & 0.0041 & 0.5926 & 1.6951 & -2.2454 & 4.08\tabularnewline
\hline 
RTC & 3 & IV & -7.1187 & 0.4480 & 1.0525 & 1.0798 & 0.4598 & 0.0088 & 0.6708 & 1.7778 & -3.7050 & 3.48\tabularnewline
\hline 
RTZ & 1 & I & -0.6989 & 0.6444 & 1.7687 & 0.9841 & 0.7326 & 8.3996 & 0.5998 & 1.5167 & -0.8796 & 1.86\tabularnewline
\hline 
RTZ & 1 & II & -3.9194 & 0.3509 & 1.0213 & 1.0540 & 0.8643 & 2.1014 & 0.2251 & 1.6100 & -1.7900 & 2.10\tabularnewline
\hline 
RTZ & 2 & I & -1.1636 & 0.8492 & 1.8549 & 0.9878 & 0.7606 & 7.2968 & 0.6346 & 1.6906 & -0.8305 & 3.06\tabularnewline
\hline 
RTZ & 2 & II & -6.9405 & 0.3777 & 1.0652 & 1.0618 & 0.9325 & 1.0444 & 0.2408 & 1.7705 & -3.8064 & 3.39\tabularnewline
\hline 
RTZ & 3 & I & -1.7681 & 0.6155 & 1.6327 & 1.0472 & 1.1094 & 4.3515 & 0.6212 & 1.6716 & -0.4518 & 3.32\tabularnewline
\hline 
RTZ & 3 & II & -6.9420 & 0.2634 & 1.0710 & 1.0583 & 0.6337 & 0.0005 & 0.5543 & 1.6732 & -3.5774 & 3.44\tabularnewline
\hline 
RTCZ & 2 & III & -6.3698 & 0.2626 & 1.1139 & 1.0794 & 0.8666 & 0.1547 & 0.5521 & 1.7007 & -2.5299 & 4.13\tabularnewline
\hline 
RTCZ & 2 & IV & -7.1531 & 0.3382 & 1.0594 & 1.0830 & 1.1192 & 0.1113 & 0.5253 & 1.6354 & -3.7004 & 3.32\tabularnewline
\hline 
RTCZ & 3 & III & -7.2104 & 0.2296 & 1.0769 & 1.0881 & 1.0376 & 0.0064 & 0.7990 & 1.5220 & -3.1565 & 4.08\tabularnewline
\hline 
RTCZ & 3 & IV & -7.0696 & 0.3242 & 1.0603 & 1.0572 & 0.8060 & 0.0103 & 0.4343 & 1.6757 & -3.7246 & 3.33\tabularnewline
\hline 
RTCZ & 4 & III & -6.8575 & 0.3248 & 1.0796 & 1.0511 & 0.8774 & 0.2604 & 0.3911 & 1.6827 & -3.2172 & 4.21\tabularnewline
\hline 
RTCZ & 4 & IV & -5.9339 & 0.1883 & 1.1421 & 1.0649 & 0.2773 & 4.35e-6 & 0.9005 & 1.6442 & -3.7037 & 3.29\tabularnewline
\hline 
RTCZ & 5 & III & -8.2560 & 0.4653 & 0.9947 & 1.0665 & 1.7552 & 1.9704 & 0.2600 & 1.5930 & -2.8293 & 4.14\tabularnewline
\hline 
RTCZ & 5 & IV & -4.0270 & 0.0326 & 1.2657 & 1.0245 & 2.4171 & 0.8344 & -0.2982 & 1.5100 & -4.5747 & 3.34\tabularnewline
\hline 
RTCZ & 6 & III & -5.6443 & 0.1462 & 1.1504 & 1.0669 & 0.8911 & 0.0042 & 0.3929 & 1.6808 & -2.5730 & 4.13\tabularnewline
\hline 
RTCZ & 6 & IV & -3.8103 & 0.0329 & 1.3015 & 1.0127 & 1.7109 & 0.4613 & -0.4411 & 1.6626 & -4.2685 & 3.41\tabularnewline
\hline 
\end{tabular}
\end{table*}

\subsection{Order Estimation Accuracy and Complexity\label{sec:Order-Estimation-Accuracy}}

Table \ref{tab:parameters} gives all coefficients $c$ and $a_{k}$,
$b_{k}$, $k=1,2,3,4$, for the different NBs and equalizer cases
as well as the corresponding maximal estimation error, $\varepsilon_{\text{max}}$,
which is the maximal value of $|N_{\text{est}}(B,\delta)-N_{\text{min}}(B,\delta)|$
over all values of $B$ and $\delta$. As seen, $\varepsilon_{\text{max}}\leq4.26$
in all cases. Further, as seen in the second term in \eqref{eq:Nest}
(which dominates the estimate), since $a_{4}\approx1$, the order
is roughly inversely proportional to the don't-care band $\pi-B$,
which is typical for linear-phase FIR filters \cite{Ichige2000,Lai_2008,Sheikh2011,Wang_2023,Wang_2019}.
This means that the order increases rapidly when the bandwidth approaches
a whole NB, as illustrated in Fig. \ref{fig: NB_3_mode_Mix_filtertype_4_SR2-1}.
Hence, one should preferably avoid bandwidths over some
80\% of a NB, and to enable low orders, the bandwidth needs to be below
some 50\%. A narrower bandwidth (more oversampling) also decreases
the requirements and complexity of the analog filters that follow
the DAC \cite{Johansson_14_0}.

\subsection{Comparisons Between Different Filter Types and NBs\label{sec:Complexity-Comparisons}}

\begin{figure}[t]
\centering 
 \includegraphics[scale=0.52]{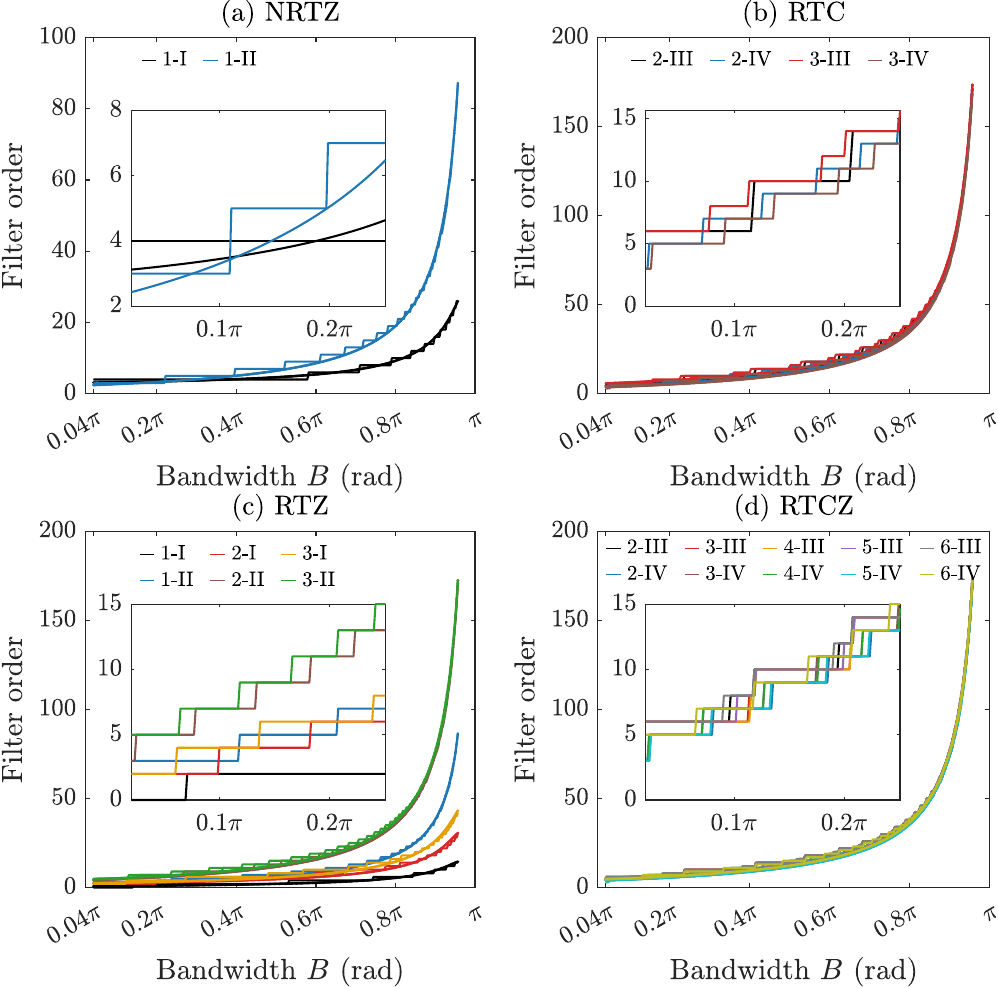} 
 \caption{Filter orders for all cases when $\delta=0.001$. Here, 1-I stands
for $NB=1$ and filter Type I, etc. Subplot (a) shows both the estimated
and minimal orders (illustrating the accuracy), whereas the other subplots
only show the minimal orders for visibility reasons.}
 \label{fig: NB_3_mode_Mix_filtertype_4_SR2-1}
\end{figure}

Figure \ref{fig: NB_3_mode_Mix_filtertype_4_SR2-1} plots the filter
orders of the different equalizers for $\delta=0.001$ which illustrates
the following features. Firstly, for the same bandwidth and equalization
accuracy, the order is lower in the first NB. This is because the
signal band starts from zero in the first NB but is centered in the
middle in the higher NBs. This implies that, for the same bandwidth,
the don't-care band is two times wider in the first NB compared to
the two don't-care bands in the higher NBs. This leads to, approximately,
a two times lower filter order in the first NB. Secondly, the order of the Type I filter is lower than those of the
Type II--IV filters. This is because the former does not have any
structural zero (see Table \ref{tab:Filter properties}). Hence, from
the equalizer complexity point of view, one should use a Type I equalizer
whenever possible, i.e., in the first three NBs. For example, in the
second NB for a bandwidth of 80\%, the orders of Type I--IV filters
are 12, 37, 38, and 37, respectively. However, as mentioned in Section
\ref{sec:DAC-Pulses-and}, in the higher NBs ($NB>1$), RTC or RTCZ
may be preferred for other reasons (e.g. lower out-of-band gains)
which means that Type III or Type IV filters should be used. Finally,
for $NB>1$, the orders of the Type II--IV filters are approximately
the same for the same bandwidth and equalization accuracy.

\section{Conclusion\label{sec:Conclusion}}

This letter considered the design and properties of linear-phase FIR
filters for the equalization of the frequency responses of four DAC
pulses that are used for DACs in multiple NBs (up to the sixth NB).
To this end, filter-order estimates, as functions of the bandwidth
and equalization accuracy, were proposed. A large set of design examples
demonstrated the accuracy of the proposed estimates.

\bibliographystyle{IEEEtran}
\bibliography{IEEEabrv.bib,references/mybib.bib}
 
\end{document}